\documentclass[prb,aps,reprint,twocolumn,superscriptaddress]{revtex4-1}

\usepackage{color}
\usepackage{graphicx}
\usepackage{amsmath}
\usepackage{txfonts}
\begin{document}

\title{Relativistic Internally Contracted Multireference Electron Correlation Methods}
\author{Toru Shiozaki}
\email{shiozaki@northwestern.edu}
\affiliation{\mbox{Department of Chemistry, Northwestern University, 2145 Sheridan Rd., Evanston, IL 60208, USA.}}
\author{Wataru Mizukami}
\affiliation{Department of Energy and Material Sciences, Faculty of Engineering Sciences, Kyushu University, 6-1 Kasuga-Park, Fukuoka, 816-8580, Japan}
\date{\today}

\begin{abstract}
We report internally contracted relativistic multireference configuration interaction (ic-MRCI), complete active space second-order perturbation
(CASPT2), and strongly contracted $n$-electron valence state perturbation theory (NEVPT2)
on the basis of the four-component Dirac Hamiltonian,
enabling accurate simulations of relativistic, quasi-degenerate electronic structure of molecules
containing transition-metal and heavy elements.
Our derivation and implementation of ic-MRCI and CASPT2 are based on an automatic code generator that translates second-quantized ans\"atze to tensor-based equations, and to efficient computer code.
NEVPT2 is derived and implemented manually.
The rovibrational transition energies and absorption spectra of HI and TlH are presented to demonstrate the accuracy of these methods.
\end{abstract}

\maketitle

\section{Introduction}
There are continued interests in accurate modeling of gas-phase thermochemistry and dynamics that involve transition-metal and heavier elements, where
relativistic effects play an important role.
For instance, scientists at the US Air Force recently performed an experiment, aiming to use chemi-ionization involving lanthanide atoms to 
alter the electron density in the ionosphere for radio-frequency communication,\cite{Shuman2015CR,Cox2015JCP}
for which accurate simulations could help analyze the experimental observation.
Another example is the reaction of $\mathrm{FeO}^+$ with a hydrogen molecule, a model reaction system for the so-called two-state reactivity,\cite{Schroder2000ACR} of which
accurate modeling still remains a challenge.\cite{Ard2014JPCA}
The spin barriers in the two-state reactivity mechanism are also ubiquitous in organometallic chemistry.\cite{Harvey2007PCCP}
Understanding these problems requires accurate description of strongly relativistic, quasi-degenerate electronic structure. 
There have been, however, only a handful of theory developments to address this challenge.\cite{Malmqvist2002CPL,Abe2006JCP,Fleig2007TCA,Fleig2012CP,Kim2014JCP} 

As a first step toward realizing predictive simulations of such processes,
we develop in this work novel computational tools that combine the four-component relativistic Dirac formalism\cite{Reiherbook}
and internally contracted multireference electron correlation methods.
Our approach is based on the four-component Dirac equation for electrons,
\begin{align}
\hat{H} = \sum_i \left[c^2 (\beta - I_4) + c(\boldsymbol{\alpha} \cdot \hat{\mathbf{p}}_i) - \sum_A^{\mathrm{atoms}} \frac{Z_A}{r_{iA}} \right] +  \sum_{i<j} \hat{g}(i,j), \label{hamil}
\end{align}
where $\boldsymbol{\alpha}$ and $\beta$ are Dirac's matrices, $\hat{g}(i,j)$ is a two-electron operator, and $c$ is the speed of light.
$Z_A$ is a charge of a nucleus $A$ (note, however, that we use finite-nucleus models in practice).
Hereafter atomic units are used unless otherwise stated.
In this work, we use the full Breit operator for electron--electron interactions, i.e.,
\begin{align}
\hat{g}(i,j) = \frac{1}{r_{ij}} - \frac{1}{2}\frac{\boldsymbol{\alpha}_i\cdot \boldsymbol{\alpha}_j}{r_{ij}} - \frac{1}{2}\frac{(\boldsymbol{\alpha}_i\cdot \mathbf{r}_{ij}) (\boldsymbol{\alpha}_j\cdot \mathbf{r}_{ij})}{r^3_{ij}}.
\end{align}
The reader may consult Ref.~\onlinecite{Shiozaki2013JCP} for details on integral evaluation associated with this operator over Gaussian basis functions.
We first perform complete active space self-consistent field (CASSCF) calculations using this Hamiltonian,\cite{Jensen1996JCP,Bates2015JCP} in which orbitals are optimized using the minimax principle, and project out the space spanned by the `negative-energy' orbitals,
a procedure called no-pair projection.\cite{Reiherbook}
An efficient Dirac-CASSCF algorithm that we have developed can be found in Refs.~\onlinecite{Bates2015JCP} and \onlinecite{Kelley2013JCP}.
After the no-pair projection procedure, the Hamiltonian in the second quantization becomes 
\begin{align}
\hat{H}_\mathrm{NP} = \sum_{xy} h_{xy} \hat{E}_{xy} +  \frac{1}{2}\sum_{xyzw} v_{xy,zw} \hat{E}_{xy,zw},
\label{mohamil}
\end{align}
where $x$, $y$, $z$, and $w$ label any electronic molecular spin orbitals (MO), and
$h_{xy}$ and $v_{xy,zw}$ are the (complex-valued) Hamiltonian matrix elements in the MO basis in chemists' notation.
$\hat{E}_{xy}$ and $\hat{E}_{xy,zw}$ are operators defined as
\begin{subequations}
\begin{align}
&\hat{E}_{xy}= a^\dagger_x a_y,\\
&\hat{E}_{xy,zw}= a^\dagger_x a^\dagger_z a_w a_y.
\end{align}
\end{subequations}
Since the MO Hamiltonian [Eq.~\eqref{mohamil}] is isomorphic to the non-relativistic counterpart (and all the eigenstates are minima in the parameter space
after the no-pair projection procedure),
standard electron-correlation methods, such as internally contracted multireference configuration interaction (ic-MRCI),\cite{Werner1982JCP,Werner1988JCP,Sham2011JCP,Saitow2013JCP} can be used in conjunction with this Hamiltonian. 
We note in passing that, even though our numerical results are based on the four-component formalism [Eq.~\eqref{hamil}], the multireference theory and programs developed in this work are equally
applicable to any two-component relativistic Hamiltonians.\cite{Liu2010MP,Saue2011CPC,Nakajima2012CR} 

In the non-relativistic framework, the ic-MRCI method has been pioneered by Werner and co-workers.\cite{Werner1982JCP,Werner1988JCP,Sham2011JCP}
The ability of ic-MRCI to accurately and consistently describe the potential energy surfaces of small-molecule reactions
has been the key to understanding many of the gas-phase reactions studied in the past decades (for instance, see Refs.~\onlinecite{Manolopoulos1993S,Alexander2002S,Wu2004S}).
Very recently ic-MRCI has been extended to incorporate density matrix renormalization group reference functions with more than 20 orbitals in the active space by Saitow et al.\cite{Saitow2013JCP} 
There are also parallel implementations of uncontracted MRCI,\cite{Lischka2011WIREs} though its computational cost is generally higher than that of ic-MRCI.

Another class of popular multireference approaches in non-relativistic theory is based on perturbation theory.
Among others the complete active space second-order perturbation (CASPT2) method\cite{Andersson1990JPC,Andersson1992JCP,Aquilante2008JCTC} is an internally contracted, multireference generalization of the standard M\o ller--Plesset perturbation theory and has been
applied to a wide variety of chemical problems.\cite{Pulay2011IJQC}
The $n$-electron valence state perturbation theory (NEVPT2)\cite{Angeli2001JCP,Angeli2002JCP} proposed by Angeli~et~al. (especially its strongly correlated variant)
uses a different zeroth-order Hamiltonian and has desirable properties such as strict size extensivity and numerical robustness against so-called intruder-state problems. 

Here we report the theory and algorithms for relativistic ic-MRCI, CASPT2, and NEVPT2 based on the four-component Dirac Hamiltonians.
This work realizes relativistic ic-MRCI and NEVPT2 for the first time, whereas
CASPT2 has been reported in the past by Abe et al.\cite{Abe2006JCP} and by Kim et al.\cite{Kim2014JCP}
The implementations of ic-MRCI and CASPT2 are facilitated by an automatic code generator, {\sc smith3}.\cite{MacLeod2015JCP,smith}
The {\sc smith3} program was previously used to derive and implement nuclear energy gradients for fully internally contracted CASPT2\cite{MacLeod2015JCP}
and has been extended in this work to incorporate equations with spin orbitals in complex arithmetic. 
Note that the automatic code generation approach has been used for relativistic single-reference coupled-cluster methods
by Hirata et al.\cite{Hirata2007JCP} and by Nataraj et al.\cite{Nataraj2010JCP} 
The generated code and the code generator are both publicly available.\cite{bagel,smith}
The NEVPT2 code is manually implemented.
In the following we sketch the outline of the theories and implementations.

\section{Theory}
\subsection{Relativistic MRCI with internal contraction}
Our ic-MRCI implementation uses fully internally contracted basis functions,
which are similar to those used in the CASPT2 theory by Roos and co-workers.\cite{Andersson1992JCP}
The correlated wave functions are parameterized as
\begin{align}
|\Psi\rangle = T_\mathrm{ref} |\Phi_{\mathrm{ref}}\rangle
+ \sum_\Omega T_\Omega \hat{E}_\Omega |\Phi_{\mathrm{ref}}\rangle, \label{param}
\end{align}
in which $T$'s are the unknown amplitudes to be determined,
$\Omega$ denotes excitation manifolds in ic-MRCI,
and $\hat{E}_\Omega$ are associated excitation operators:
\begin{align}
\hat{E}_\Omega = &\left\{\hat{E}_{ai,bj},\,  \hat{E}_{ar,bi},\,  \hat{E}_{ar,bs},\,  \hat{E}_{ai,rj},\right.\nonumber\\
&\,\left.\hat{E}_{ri,sj},\,  \hat{E}_{ar,st},\,  \hat{E}_{ri,st},\,  \hat{E}_{ai,rs}\right\}.
\label{class}
\end{align}
Hereafter $i$ and $j$ label closed orbitals, $r$, $s$, and $t$ label active orbitals, and $a$ and $b$ label virtual orbitals. 
Note that, because spin orbitals are used, $\hat{E}_{ai,rs}$ and $\hat{E}_{as,ri}$ that are distinguished in non-relativistic theories
generate identical sets of excited configurations.
The Kramers symmetry is not utilized in our ic-MRCI implementation except for integral compression.
$|\Phi_{\mathrm{ref}}\rangle$ is a relativistic multi-determinant reference function, 
\begin{align}
|\Phi_{\mathrm{ref}}\rangle = \sum_{n_++n_- = n} C^{n_+,n_-} |I^{n_+,n_-}\rangle,
\end{align}
where $n_+$ and $n_-$ are the numbers of electrons that belong to Kramers $+$ and $-$ spin orbitals,
and $n$ is the total number of active electrons.\cite{Jensen1996JCP,Bates2015JCP}

In the ic-MRCI method, the Dirac Hamiltonian is diagonalized
in the space spanned by the parameters in Eq.~\eqref{param}, i.e.,
\begin{align}
&E = \min \left[\langle \Psi|\hat{H}_\mathrm{NP}|\Psi\rangle\right],
\end{align}
under a normalization constraint.
The following $\sigma$ and $\pi$ vectors are computed from each trial vector $\psi_P$ in the same basis,
\begin{subequations}
\begin{align}
\label{eqbegin}
&(\sigma_P)_\Omega = \langle \Phi_{\mathrm{ref}}| \hat{E}_\Omega^\dagger\hat{H}_\mathrm{NP} |\psi_P \rangle,\\
&(\sigma_P)_\mathrm{ref} = \langle \Phi_{\mathrm{ref}}| \hat{H}_\mathrm{NP} |\psi_P \rangle,\\
&(\pi_P)_\Omega = \langle \Phi_{\mathrm{ref}}| \hat{E}_\Omega^\dagger |\psi_P \rangle,\\
&(\pi_P)_\mathrm{ref} = \langle \Phi_{\mathrm{ref}}| \psi_P \rangle.
\end{align}
\end{subequations}
Note that we eliminate five-particle reduced density matrices from the equations by means of
a well-known commutator trick, i.e., (using $\hat{T}_\Omega \equiv T_\Omega\hat{E}_\Omega$)
\begin{align}
&\langle \Phi_{\mathrm{ref}}| \hat{E}_{\Omega'}^\dagger\hat{H}_\mathrm{NP} \hat{T}_\Omega |\Phi_{\mathrm{ref}}\rangle \nonumber\\
& \quad = \langle \Phi_{\mathrm{ref}}| \hat{E}_{\Omega'}^\dagger [\hat{H}_\mathrm{NP}, \hat{T}_\Omega] |\Phi_{\mathrm{ref}}\rangle
+ \langle \Phi_{\mathrm{ref}}| \hat{E}_{\Omega'}^\dagger \hat{T}_\Omega | \Phi_{\mathrm{ref}}\rangle E_{\mathrm{ref}},\label{eqend}
\end{align}
where $\Omega$ and $\Omega'$ belong to the same excitation class in Eq.~\eqref{class}.
A Hamiltonian matrix is then constructed within the subspace spanned by the trial vectors,\cite{Davidson1975JCompP}
\begin{align}
H_{PQ} = \mathbf{T}_P^\dagger \boldsymbol{\sigma}_Q, \quad S_{PQ} =  \mathbf{T}_P^\dagger \boldsymbol{\pi}_Q, 
\end{align}
and diagonalized to obtain the coefficients ($c_P$) that constitute an optimal linear combination of the trial vectors:
\begin{align}
\sum_Q H_{PQ} c_Q = E\sum_Q S_{PQ} c_Q. 
\end{align}
Using these quantities, the residual vectors are
\begin{align}
\mathbf{R} = \sum_P c_P \left[\boldsymbol{\sigma}_P - E \boldsymbol{\pi}_P \right], 
\end{align}
from which we generate a new set of trial vectors (see below).

The working equations [Eqs.~\eqref{eqbegin}--\eqref{eqend}] for $\sigma$-vector formation can be expressed in terms of reduced density matrices;
therefore, it is essentially identical to the non-relativistic counterpart except for spin symmetry in the latter.
The explicit formulas consist of ca.~750 tasks, most of which are tensor contractions. They can be found in supporting information.\cite{supp}
The equations were implemented into efficient computer code using the automatic code generator {\sc smith3}.\cite{MacLeod2015JCP,smith}
First, {\sc smith3} performs Wick's theorem to convert second-quantized expressions to a list of diagrams represented by tensors and their contractions.
Next it factorizes the diagrams to a tree of binary tensor contractions. Finally the tree is translated to computer code that is compiled and linked to the {\sc bagel} package.\cite{bagel}
See Refs.~\onlinecite{Hirata2003JPCA,Hirata2006TCA,Shiozaki2008PCCP} for further information on automatic code generation.

At the end of each ic-MRCI calculation, the Davidson correction is added to the total energy to approximately account for
size-extensivity errors.\cite{Langhoff1974IJQC}
The correction is 
\begin{align}
\Delta E_{+\mathrm{Q}} = \left(\frac{1-T_\mathrm{ref}^2}{T_\mathrm{ref}^2}\right) E_\mathrm{corr},
\end{align} 
where $T_\mathrm{ref}$ is the weight of the reference configuration in the correlated wave function [see Eq.~\eqref{param}],
and $E_\mathrm{corr}$ is the correlation energy from ic-MRCI calculations.

\subsection{Relativistic CASPT2 and NEVPT2}
The second-order perturbation methods, CASPT2 and NEVPT2, are defined as minimization of the so-called Hylleraas functional,
\begin{align}
E = \min\left[\langle \Psi^{(1)} | \hat{H}^{(0)} - E^{(0)} | \Psi^{(1)} \rangle + 2\Re\langle  \Psi^{(1)} | \hat{H}_\mathrm{NP} | \Phi_{\mathrm{ref}} \rangle\right].
\end{align} 
In CASPT2, the zeroth-order Hamiltonian $\hat{H}^{(0)}$ is chosen to be a projected Fock operator
\begin{align}
 \hat{H}^{(0)} = \hat{P}\hat{f}\hat{P} + \hat{Q}\hat{f}\hat{Q},
\end{align}
where $\hat{P}$ is a projector to the reference configuration and $\hat{Q}$ is its orthogonal compliment.
The first-order wave function $\Psi^{(1)}$ is parameterized as in Eq.~\eqref{param}. 
The minimization is performed by solving a set of linear equations using a subspace algorithm.
The construction of residual vectors,
\begin{align}
R_\Omega = 2\left[\langle \Omega |  \hat{H}^{(0)} - E^{(0)} |\psi_p\rangle +  \langle \Omega | \hat{H}_\mathrm{NP} | \Phi_{\mathrm{ref}} \rangle\right], 
\end{align}
is akin to  (but simpler than) that in ic-MRCI.
Here we used $\langle \Omega|\equiv \langle \Phi_\mathrm{ref}|\hat{E}_\Omega^\dagger$.
For details on the relativistic CASPT2 equations, see earlier reports by Abe et al.\cite{Abe2006JCP} and Kim et al.\cite{Kim2014JCP} 

In NEVPT2, the zeroth-order Hamiltonian is defined using Dyall's Hamiltonian\cite{Dyall1995JCP} as 
\begin{align}
\hat{H}^{(0)} = \hat{P}\hat{H}_\mathrm{NP}\hat{P} + \sum_{\omega}|\Phi_\omega\rangle E_\omega  \langle\Phi_\omega|,  
\label{nevh0}
\end{align}
where $\omega$ is the excitation class in Eq.~\eqref{class} and $\Phi_\omega$ is defined as
\begin{align}
|\Phi_\omega\rangle = \frac{\hat{P}_\omega \hat{H}_\mathrm{NP}  |\Phi_\mathrm{ref}\rangle}{ \sqrt{\langle \Phi_\mathrm{ref}|\hat{H}_\mathrm{NP}  \hat{P}_\omega \hat{H}_\mathrm{NP}  |\Phi_\mathrm{ref}\rangle}}.
\end{align}
$\hat{P}_\omega$ is a projector onto $\omega$, and the denominator accounts for normalization.
$E_\omega$ that appears in Eq.~\eqref{nevh0} is
\begin{align}
E_\omega = \langle \Phi_\omega| \hat{H}_\mathrm{NP} | \Phi_\omega\rangle. 
\end{align}
The wave function is parameterized using the so-called strong contraction scheme, i.e.,
\begin{align}
|\Psi\rangle = T_\mathrm{ref}|\Phi_\mathrm{ref}\rangle + \sum_\omega T_\omega |\Phi_\omega\rangle.
\end{align} 
Since $\hat{H}^{(0)}$ of NEVPT2 does not include off-diagonal couplings between different $\omega$, the equations can be solved
without iterative procedures.
The working equations for relativistic NEVPT2 can be obtained
by dropping the factors of 2 that stem from spin summations in the non-relativistic equations in Ref.~\onlinecite{Angeli2002JCP}.
The explicit formulas are provided in supporting information.\cite{supp}

\subsection{Wave function updates in ic-MRCI and CASPT2}
Internally contracted basis functions ($\hat{E}_\Omega |\Phi_\mathrm{ref}\rangle$)
are not orthogonal with each other and sometimes linearly dependent;\cite{Werner1988JCP} 
therefore, one has to take into account the overlap matrix when updating the amplitudes.
The generation of trial vectors is performed as the following.
Let us consider as an example the amplitudes associated with $\hat{E}_{ar,bs}$. In this case, the overlap and (approximate) diagonal Hamiltonian matrix elements,
$\mathbf{S}$ and $\mathbf{F}$, respectively, are 
\begin{subequations}
\label{smat}
\begin{align}
&S_{rs,r's'} = \langle \Phi_{\mathrm{ref}}|\hat{E}_{rr',ss'}|\Phi_{\mathrm{ref}}\rangle,\\
&F_{rs,r's'} = \sum_{tt'}\langle \Phi_{\mathrm{ref}}|\hat{E}_{rr',ss',tt'}|\Phi_{\mathrm{ref}}\rangle f_{tt'},
\end{align}
\end{subequations}
where $\hat{E}_{rr',ss',tt'} = a^\dagger_r\hat{E}_{ss',tt'}a_{r'}$.
We calculate $\mathbf{S}^{-1/2}$ while projecting out the linearly dependent part so that $(\mathbf{S}^{-1/2})^\dagger \mathbf{S} \mathbf{S}^{-1/2}$ is a unit matrix (the eigenvalues that are smaller than $1.0\times 10^{-8}$ are discarded),
which is then used to form 
\begin{align}
\tilde{\mathbf{F}} = (\mathbf{S}^{-1/2})^\dagger \mathbf{F} \mathbf{S}^{-1/2}.
\end{align}
Next $\tilde{\mathbf{F}}$ is diagonalized to yield a transformation matrix $\mathbf{U}$,
\begin{align}
\tilde{\mathbf{F}} = \mathbf{U} {\boldsymbol{\lambda}} \mathbf{U}^\dagger,
\label{fdiag}
\end{align}
with a diagonal matrix ${\boldsymbol{\lambda}}$.
Defining $\mathbf{X} = \mathbf{U}^\dagger \mathbf{S}^{-1/2}$,
we arrive at the formula for generating new trial vectors from residual vectors:
\begin{align}
(\psi_{p+1})_{ar,bs} = \sum_D \left[ \sum_{r's'}\frac{R_{ar',bs'} X_{D,r's'}}{E^{(0)} - \lambda_D - \epsilon_a - \epsilon_b}\right]X^\ast_{D,rs},
\end{align}
where $\epsilon_a$ is an orbital energy (i.e., $\epsilon_a = f_{aa}$) and $D$ labels the eigenvalues in Eq.~\eqref{fdiag}, the number of which is equal to or smaller than the numbers of rows and columns of the overlap matrix [Eq.~\eqref{smat}]. 
This formula implies that in ic-MRCI updates
the inverse of $\hat{H}_\mathrm{NP}-E$ is approximated by that of the diagonal part of the CASPT2 equation.\cite{Andersson1990JPC}

\subsection{Computation of rovibrational spectra}
Rovibrational energy levels of diatomic molecules in their $\Sigma$ states can be calculated by solving an effective one-dimensional Schr{\"o}dinger equation (in this section we avoid use of atomic units for clarity),
\begin{align}
\left[-\frac{\hbar^2}{2\mu}\frac{{\rm d}^2}{{\rm d}r^2}
+ V(r) + \frac{\hbar^2}{2\mu r^2}J(J+1)\right]
\Psi_{\nu,J}\left(r\right)
= E_{\nu,J} \Psi_{\nu,J}\left(r\right),
\label{radse}
\end{align}
in which $\nu$ and $J$ are the vibrational and rotational quantum numbers, respectively,
and $\mu$ is the reduced mass. The third term of the Hamiltonian accounts for the Coriolis coupling.
The rotation--vibration coupling is, therefore, variationally included in the calculations.

The line intensity $I_{\tilde{\nu}}$ associated with the transition energy $\tilde{\nu}$ can be computed as\cite{Bernath2005Spectra}
\begin{align}
  I_{\tilde{\nu}} = \frac{(2J_f+1)}{8\pi c Q \tilde{\nu}^2}
  \mathcal{A}_{\nu_i,J_i\to \nu_f,J_f}
  e^{-{E_i}/{kT}}
  \left(
  1-e^{-hc\tilde{\nu}/kT}
  \right),
\end{align}
in which $E_{i}$ is the energy of the initial state and $k$ is the Boltzmann constant.
The partition function $Q$ at a temperature $T$ is evaluated using
\begin{align}
  Q = \sum_{l} \left(2J_l+1\right) e^{-{E_l}/{kT}},
\end{align}
where $l$ runs over rovibrational states.
We used  $T=296$~K.
The quantum numbers of initial (final) states are labeled by $\nu_i$ and $J_i$ ($\nu_f$ and $J_f$).
Using the rovibrational wave functions ($\Psi_{\nu_i,J_i}$ and $\Psi_{\nu_f,J_f}$) and the dipole-moment function $M(r)$,
the Einstein coefficient $\mathcal{A}_{\nu_i,J_i\to \nu_f,J_f}$ is
\begin{align}
  \mathcal{A}_{\nu_i,J_i\to \nu_f,J_f}
  = \frac{8\pi^2\tilde{\nu}^3}{3\epsilon_{0}c^3\hbar}
  \frac{S_{J_i,J_f}}{2J_i+1}
  \left| \langle \Psi_{\nu_i,J_i}|M(r)|\Psi_{\nu_f,J_f} \rangle \right|^2,
\label{defA}
\end{align}
where $\epsilon_{0}$ is the vacuum permittivity
and $S_{J_i,J_f}$ is the H{\"o}nl--London factor,\cite{Hansson2005}
which is ${\rm{max}}(J_i,J_f)$ for the electronic ground states of HI and TlH.

\begin{table}
\caption{Root-mean-square deviations of the rovibrational transition energies of H$^{127}$I and $^{205}$TlH in $\rm{cm^{-1}}$ computed by the four-component methods.
The HITRAN database\cite{Hitran2012JQSRT} and experimental data\cite{Urban1989CPL} were used as references.\label{rovib}}
\begin{ruledtabular}
  \begin{tabular}{lccccc}
 & \multicolumn{1}{c}{CASSCF} & \multicolumn{1}{c}{CASPT2} & \multicolumn{1}{c}{NEVPT2} & \multicolumn{1}{c}{MRCI+Q} &  \multicolumn{1}{c}{Origin}\\\hline
HI\\
$\nu=0\to1$ & 120 &  36 & 21 &  8 & 2230 \\
$\nu=0\to2$ & 245 &  73 & 43 & 15 & 4379 \\
$\nu=0\to3$ & 378 & 117 & 68 & 24 & 6448 \\
$\nu=0\to4$ & 519 & 163 & 96 & 34 & 8435 \\
TlH\\
$\nu=0\to1$ & 92 & 34 & 47 & 17 & 1345 \\
$\nu=1\to2$ & 93 & 33 & 46 & 15 & 1300 \\
$\nu=2\to3$ & 92 & 33 & 46 & 13 & 1255 \\
\end{tabular}
\end{ruledtabular}
\end{table}

\section{Numerical Results}
\begin{figure}
\includegraphics[keepaspectratio,width=0.48\textwidth]{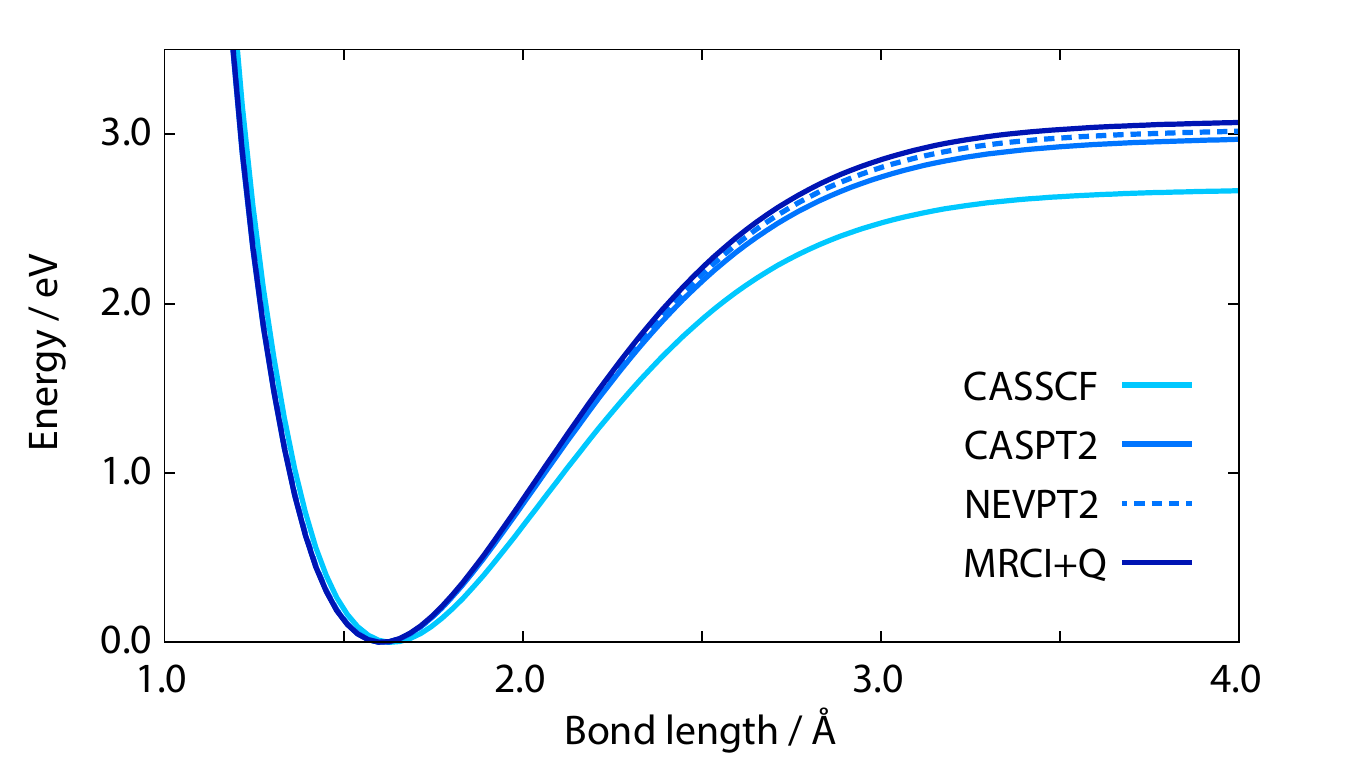}
\caption{Potential energy curves of HI computed by four-component CASSCF, CASPT2, NEVPT2, and ic-MRCI+Q. The experimental
bond length and dissociation energy are 1.609~{\AA} and 3.20~eV, respectively.\label{hipec}}
\end{figure}

\begin{figure*}[t]
\includegraphics[keepaspectratio,width=0.95\textwidth]{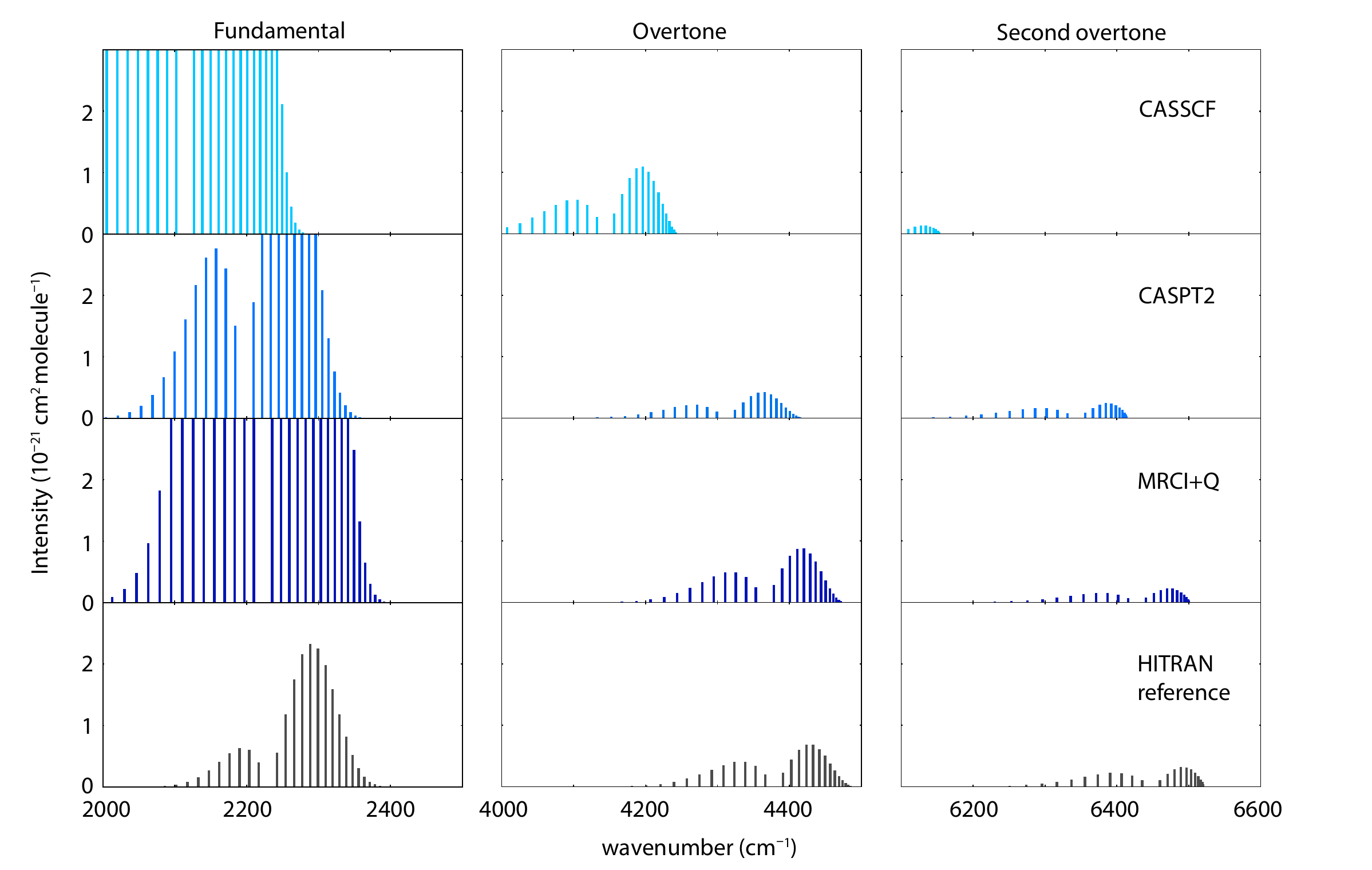}
\caption{Simulated rovibrational absorption spectra of H$^{127}$I
  at 296K using four-component CASSCF, CASPT2, and ic-MRCI+Q.
  The bottom panels are the observed lines from the HITRAN database
  (hyperfine-split lines are averaged for comparison).
  \label{hispec}}
\end{figure*}

First, to benchmark the accuracy,
we applied four-component CASSCF, CASPT2, NEVPT2, and ic-MRCI+Q to an HI molecule, for which there are reliable experimental reference data.\cite{Hitran2012JQSRT}
Uncontracted Dyall's cv3z\cite{Dyall2006TCA} and uncontracted cc-pVTZ\cite{Dunning1989JCP} basis sets were used for I and H, respectively.
Gaussian-type nuclear charge distributions were used.\cite{Visscher1997ADNDT}
The $4s$, $4p$, $4d$, $5s$, and $5p$ electrons of I and the $1s$ electron of H were correlated (i.e., 26 correlated electrons; 28 electrons were frozen),
among which $5s$, $5p$ of I and $1s$ of H were treated in the active space.
In correlated calculations, virtual orbitals were truncated at 55~$E_\mathrm{h}$.
The total number of correlated spin orbitals was 206. 
The computed potential energy curves relative to their minima are shown in Fig.~\ref{hipec}.
The equilibrium bond lengths obtained by CASPT2, NEVPT2, and ic-MRCI+Q were 1.608, 1.609, and 1.606~\AA{}, respectively, which are in good agreement with
the experimental value (1.609~\AA{}).\cite{Herzbergbook}
The dissociation energies $D_e$ were estimated via extrapolation to be 3.0, 3.0, and 3.1~eV, respectively.
The experimental value is 3.20~eV.\cite{Herzbergbook}

We then simulated the absorption spectra based on
these potential energy curves interpolated by five-point piece-wise polynomials.
Dipole moments were computed at each point as electric-field derivatives [$M(r) = \partial E(r)/\partial \mathcal{E}_z$ where $\mathcal{E}_z$ is an external electric field along the molecular axis] using finite difference formulas.
The Level 8.2 program\cite{level8.2} was used to solve the radial Schr{\"o}dinger equation [Eq.~\eqref{radse}] and to evaluate $\mathcal{A}_{\nu_i,J_i\to \nu_f,J_f}$ [Eq.~\eqref{defA}].
The partition function and absorption spectra were computed using a program of Yorke et al.\cite{pn_exomol}
The computed spectra for the fundamental, overtone, and second overtone transitions are presented in Fig.~\ref{hispec},
in which the HITRAN reference spectra\cite{Hitran2012JQSRT} are also shown. 
Overall, the line positions were accurately reproduced by ic-MRCI+Q within 0.5~\% (8~cm$^{-1}$ for the fundamental transitions and 34~cm$^{-1}$ for
the third overtone transitions),
attesting to the consistent accuracy of ic-MRCI+Q throughout potential energy surfaces;
The line intensity of the overtone and second overtones agreed well.
Our results overestimated the intensity of the fundamental transitions,
which is mainly because the intensity is largely suppressed by
the almost flat dipole-moment curve around the equilibrium geometry;
therefore, it is highly sensitive to the accuracy of the computed dipole moments.\cite{Li2013JQSRT}
The errors in the line positions computed by CASPT2 and NEVPT2 were found three or four times larger than those by ic-MRCI+Q. 

\begin{figure}
\includegraphics[keepaspectratio,width=0.48\textwidth]{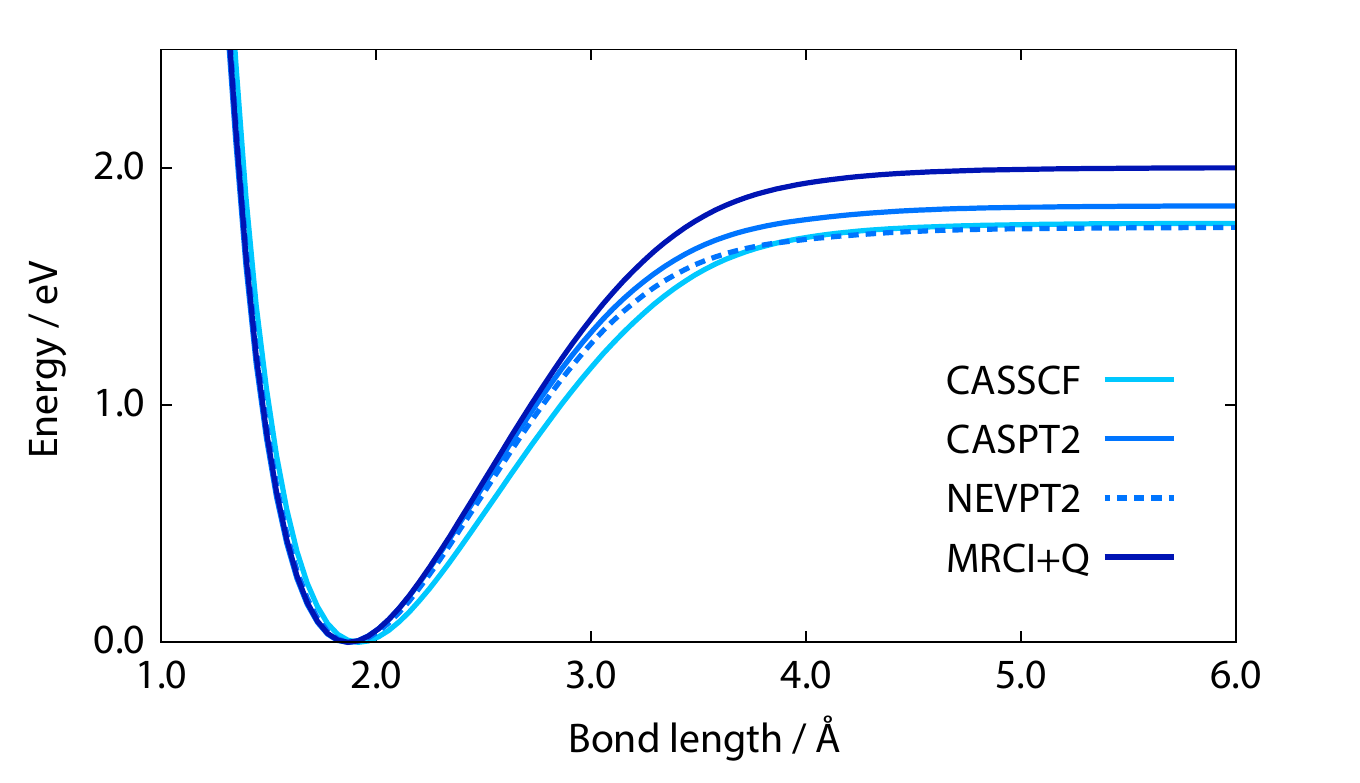}
\caption{Potential energy curves of TlH computed by four-component CASSCF, CASPT2, NEVPT2, and ic-MRCI+Q. The experimental
bond length and dissociation energy are 1.872~{\AA} and 2.06~eV, respectively.\label{tlhpec}}
\end{figure}

\begin{figure}
\includegraphics[keepaspectratio,width=0.48\textwidth]{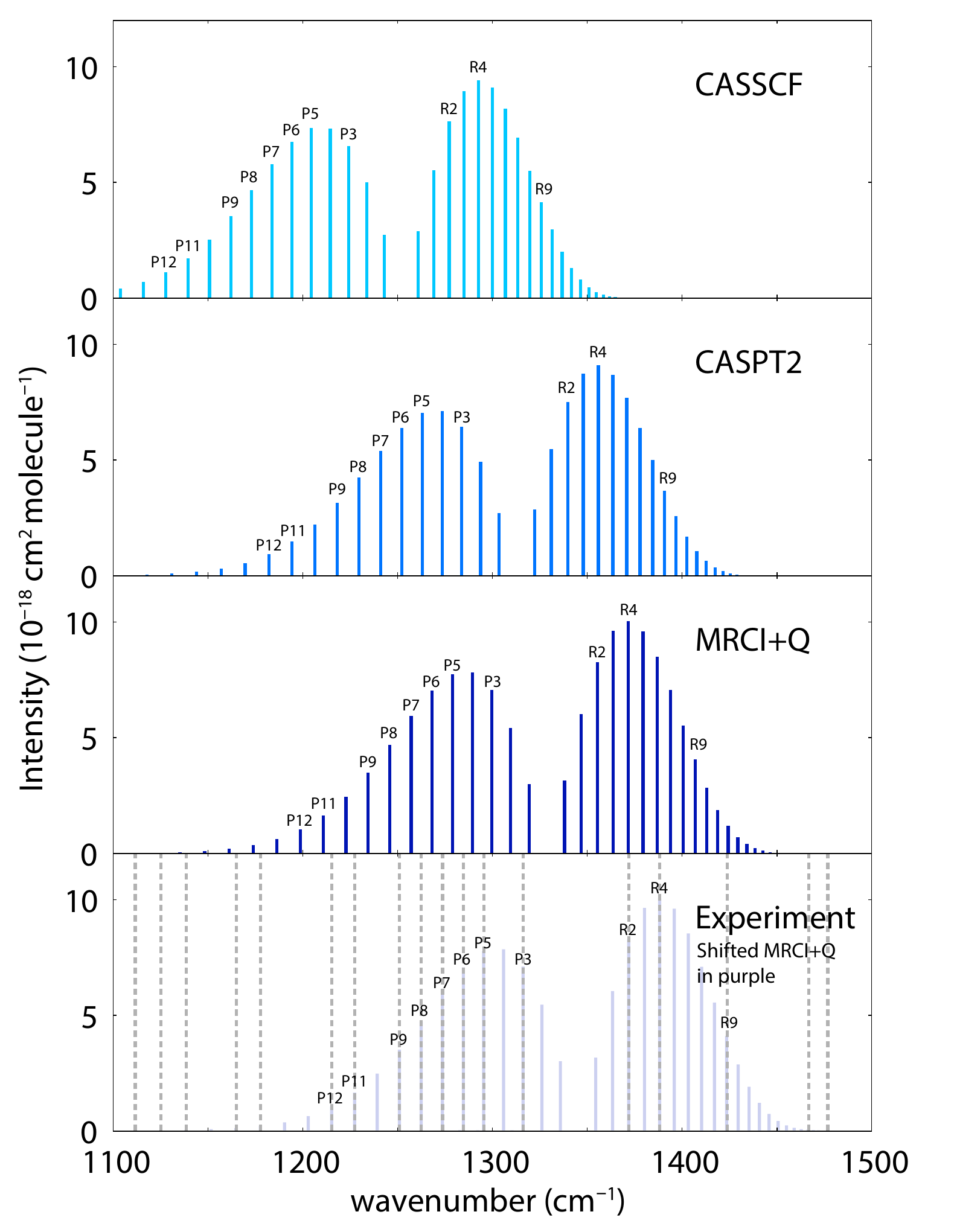}
\caption{Simulated rovibrational absorption spectra of $^{205}$TlH
  at 296K using four-component CASSCF, CASPT2, and ic-MRCI+Q.
  Dotted lines in the bottom panel are the experimental line positions taken from Ref.~\onlinecite{Urban1989CPL}
  superimposed by shifted ic-MRCI+Q spectra.
  \label{tlhspec}}
\end{figure}

Next, we calculated the potential energy curve of TlH using CASSCF, CASPT2, NEVPT2, and ic-MRCI+Q.
The electronic structure of TlH around the equilibrium geometry has been studied by many authors.\cite{Fagri2001TCA,Zeng2010JCP,Knecht2014JCP}
We used uncontracted Dyall's cv3z\cite{Dyall2006TCA} and uncontracted cc-pVTZ\cite{Dunning1989JCP} basis sets for Tl and H, respectively,
in conjunction with Gaussian-type nuclear charge distributions.\cite{Visscher1997ADNDT}
The full-valence active space (4 electrons in the $6s$ and $6p$ orbitals of Tl and the $1s$ orbital of H) was used.
The $5s$, $5p$, $4f$, $5d$, $6s$, and $6p$ electrons of Tl and the $1s$ electron of H were correlated (i.e., 36 correlated electrons).
The virtual orbitals were again truncated at 55~$E_\mathrm{h}$, resulting in 248 correlated spin orbitals.
The potential energy curves of TlH computed by four-component CASSCF, CASPT2, NEVPT2, and ic-MRCI+Q are shown in Fig.~\ref{tlhpec}.
The dissociation energy $D_e$ from ic-MRCI+Q (2.00~eV) was in excellent agreement with the experimental value ($2.06$~eV),\cite{Herzbergbook} while
CASPT2 underestimated it by 0.2~eV (1.84~eV).
The equilibrium bond length ($1.872$~\AA{}) was also accurately reproduced by ic-MRCI+Q ($1.872$~\AA{}). 
Those by CASPT2 and NEVPT2 were 1.870 and 1.885~\AA{}, respectively.
NEVPT2 was found less accurate than CASPT2 for this molecule, and its accuracy deteriorated as the bond is stretched.  

The absorption spectra of TlH were likewise computed using
the energies at 20 grid points between 1.3 {\AA} and 6.0 {\AA}.
The computed spectra are presented in Fig.~\ref{tlhspec}. 
The experimental line intensity was not found in the literature.
The mean-root-square errors in the computed rovibrational transition energies are also listed in Table.~\ref{rovib},
in which the experimental results from Ref.~\onlinecite{Urban1989CPL} are used as reference values.
The errors in the transition energies were around 35, 45, and 15~cm$^{-1}$ for CASPT2, NEVPT2, and ic-MRCI+Q. 
Apart from the shift, the line positions computed by ic-MRCI+Q agree perfectly with the experimental results. 
The remaining errors include incomplete treatment of dynamical correlation in the ic-MRCI+Q model, the effects of the higher-order quantum-electrodynamics interactions,
and the non-Born--Oppenheimer contributions.

The wall times for one iteration of relativistic CASPT2 and ic-MRCI on TlH were
roughly 2 and 80 minutes using two Xeon E5-2650 CPUs (2.0~GHz, 8 cores each) on a single node.
The wall time for non-relativistic ic-MRCI per iteration is about 16 seconds; therefore, relativistic ic-MRCI is roughly 300 times more expensive than the non-relativistic counterpart.
A factor of $2^6=64$ stems from the fact that relativistic ic-MRCI does not use spin symmetry. An additional factor of 3 should be ascribed to
matrix multiplication in complex arithmetic that is three times as expensive as that in real arithmetic.
The rest is due to other factors such as caching and optimized libraries.
 
\section{Conclusions}
In summary, we have developed four-component relativistic ic-MRCI, CASPT2, and NEVPT2 based on the Dirac Hamiltonian and full internal contraction.
The relativistic ic-MRCI and CASPT2 programs have been implemented using automatic code generation.
The programs are interfaced to the open-source {\sc bagel} package.\cite{bagel} The code generator {\sc smith3} is also publicly available.\cite{smith}
The accuracy of these methods has been presented by computing the entire potential energy curves of HI and TlH and directly comparing 
calculated rovibrational transition energies with the experimental data.
It has been shown that ic-MRCI+Q can reproduce experimental transition energies with 0.5~\% and 1~\% accuracy for HI and TlH, respectively,
up to high-lying rovibrational transitions using uncontracted triple-$\zeta$ basis sets without any corrections or extrapolations.

Currently the size of ic-MRCI and CASPT2 calculations is limited by the memory requirement for two-electron MO integrals that are stored in core,
which is somewhat problematic especially because uncontracted one-electron basis functions (with energy cut-offs) have to be used for heavy elements.
Furthermore, wall times for multi-state ic-MRCI calculations scale cubicly with respect to the number of states included in the calculation, which become prohibitively long when several states are included in the calculations.
To address these problems, the parallelization of the programs
based on the {\sc tiledarray} library of Calvin and Valeev\cite{tiledarray} is under development in our group.
Our relativistic NEVPT2 code does not store 4-index intermediates and is heavily parallelized (to be presented elsewhere);
therefore, it is ready for use in chemical applications.

\section*{Supporting Information}
The working equations for relativistic NEVPT2 and the rovibrational transition energies and absorption spectra
of HI and TlH can be found in supporting information. 
The computer-generated ic-MRCI equations are also included.

\begin{acknowledgments}
T.S. has been supported by the Air Force Office of Scientific Research Young Investigator Program (AFOSR Grant No.~FA9550-15-1-0031).
The development of the relativistic CASSCF program, on which this work is based, has been supported by the National Science Foundation CAREER Award (CHE-1351598).
W.M. has been supported by Grant-in-Aid for Young Scientists (B) (Grant No. 15K17815) from the Ministry of Education, Culture, Sports, Science and Technology Japan (MEXT).
\end{acknowledgments}

\end{document}